\documentclass[11pt]{article}
\usepackage[margin=1in]{geometry}
\usepackage[T1]{fontenc}
\usepackage[utf8]{inputenc}
\usepackage{lmodern}

\title{The indiscriminate adoption of AI threatens the foundations of academia}
\author{Roberto Trotta \\ SISSA \& Imperial College London}
\date{\today}

\begin{document}

\maketitle
\begin{abstract}
Artificial intelligence offers much promise, but its use in scientific research should be restrained so that the primary aim of academia --- advancing knowledge for humans --- is safeguarded.
\end{abstract}

Over a drink at the end of a scientific conference last summer a colleague at a world-leading institution enthused: ``I can now do in a day what used to take me three months!''. I was asking after the practice of `vibe coding', a recent trend that replaces traditional coding with verbal interaction with a specialized large language model (LLM) which, following the user's instructions, writes, tests and refines complex code at a pace unmatched by humans. As scientific data analysis becomes heavily reliant on artificial intelligence (AI) --- including cosmology, my own field --- this new approach promises to increase speed and efficiency manyfold. Few doubt that AI will be indispensable for processing, analysing and interpreting the vast amount of data produced by upcoming ground- and space-based observatories.

Vibe coding is only one of the AI-based practices with the potential to turbocharge research. `Agentic AI' refers to teams of AI `agents' --- instances of LLMs specializing into sub-tasks --- working in concert towards a larger goal. From the definition of an appropriate research question, to the formulation of a hypothesis, the development and testing of suitable code to analyse the data, to the writing of a full research paper complete with figures and codebase, ready for submission to a journal, AI agents aim at ultimately conducting wholly autonomous original research. Recent examples include a virtual biomedical lab to design nanobodies \cite{kudiabor2024}, a system to manage the whole observation process at a telescope \cite{wang2025}, and a research assistant capable of producing a hundred plausible-looking journal articles in any discipline in an afternoon --- one of which has been accepted at the first scientific conference featuring AI as the primary author and reviewer of all contributions.

Yet behind the excitement and sense of possibility lurk deeper issues. Scientific coding is often much more than software engineering, for it requires an understanding of the underlying physical and statistical models. The process can be automated for tasks the scientist is already familiar with, but tackling entirely novel problems requires complex chains of reasoning that remain (for now) beyond the capabilities of LLM agents. Even replication of existing research papers in astrophysics remains challenging, with the best agents scoring less than 20\% on this task \cite{ye2025}. Applying LLM agents to wholesale scientific reasoning risks producing derivative work, as it ultimately relies on concepts already present in the training data. Research into the use of generative AI for short stories concludes that it stifles collective novelty \cite{doshihauser2024}, which might equally apply to scientific creativity.

Apart from the well-known phenomenon of `hallucinations' (entirely made-up content presented as fact, a problem that stubbornly plagues even the latest LLM versions), the output of LLMs is often fundamentally unexplainable to humans, as LLMs' rationalizations of their own `chain of thought' have been shown to merely retrofit the expected explanation, not to reflect the actual reasoning process. The danger is that AI-generated scientific results become opaque due to the weakening of step-by-step reproducibility. Furthermore, errors in AI-generated output will become increasingly difficult to spot and correct in an academic world awash with what could be called `AI science slop'.

But the real risk is the impact on human scientists. Outsourcing deep engagement with scientific questions to LLM agents will reduce our own scientific sharpness. A recent comparative study \cite{kosmyna2025} highlights the decline in neural connectivity and cognitive ability of participants who were assisted by an LLM to write essays: relying on LLMs resulted in long-term neural, behavioural and linguistic underperformance. Creative thinking, too, is suppressed in controlled experiments comparing the performance of unassisted humans with that of subjects using LLMs to either provide the answer or help them think through the steps to the solution \cite{kumar2025}. This is true of both convergent and divergent thinking, appearing to suggest that depending on LLMs might impair our ability to come up with our own ideas. It seems plausible that the general cognitive trends highlighted by these studies carry over to scientific reasoning and coding.

The repercussions on the next generation of scientists might be even more severe: if the difficult, uncomfortable process of learning to reason like a scientist is replaced by a prompt to a chatbot, the critical thinking and depth of expertise that enables human scientists to supervise the output of LLM agents will be diminished. Research students might become little more than prompt engineers. Systemic changes to the role and funding of higher education and research institutions might ensue, as expensive PhD fellowships might swiftly be replaced by far cheaper API credits to run an army of AI agents on commercial platforms. Scientific literacy might wane within years, not decades.

The academic publishing system, so far based on voluntary human peer review, is already on the verge of collapse under the strain of an exponential increase in papers \cite{hanson2024}, a trend that will only intensify with AI agents-generated research: NeurIPS, one of the most prestigious machine learning conferences in the world, saw the number of submissions double from 2020 to 2025. The Association for the Advancement of AI Conference 2026 is piloting AI-assisted peer review to handle a record-breaking 31,000 submissions. Nobel prize laureate Venki Ramakrishnan believes that ``eventually these papers will all be written by an AI agent and then another AI agent will actually read them, analyse them and produce a summary for humans''.

Students, scientists and institutions might all feel trapped in an AI arms race. But uncritically hastening the advent of ``machines [that] will eventually compete with men [sic] in all purely intellectual fields'' (as AI pioneer Alan Turing hoped in 1950 \cite{turing1950}) will undermine the very reasons of academia's existence: that of advancing knowledge for humans, educating young minds and improving understanding of our place in the world. A debate involving both scientists and humanists on the role of AI-powered research is urgently needed, to ensure that the science of the future retains its essential human quality.

\end{document}